\documentclass[letterpaper]{article}

\usepackage[T1]{fontenc}

\usepackage{geometry}
\usepackage{setspace}

\usepackage[articletitle=true]{achemso}

\usepackage{graphicx}
\usepackage{float}
\newfloat{scheme}{htbp}{los}
\floatname{scheme}{Scheme}
\floatname{chart}{Chart}
\newfloat{graph}{htbp}{loh}

\usepackage{chemformula} 
\usepackage[version = 4]{mhchem} 
\usepackage{upgreek}
\usepackage{authblk}
\author[1]{Lai Chen}
\author[1]{Rui Wang}
\author[1]{Li-Gang Wang*}
\affil[1]{Zhejiang Key Laboratory of Micro-nano Quantum chips and Quantum control and School of Physics, Zhejiang University, Hangzhou 310058, China}

\title{Tornado Wave-Encoded Point Spread Functions for Multi-Period Three-Dimensional Localization Imaging}
\date{*Email: lgwang@zju.edu.cn}

\begin{document}

\maketitle

\begin{abstract}
  Three-dimensional (3D) localization microscopy using rotational point spread functions (PSFs) is fundamentally challenged by limited axial range, poor PSF compactness, and depth-decoding ambiguity—particularly in dense specimens with heavy PSF overlap. To address these limitations, we introduce a 3D localization approach based on tornado wave (ToW)-PSFs, combined with a convolutional neural network (CNN) designed to decode continuous rotational patterns and resolve multi-period axial ambiguity. Our engineered PSF encodes the 3D position of each nanoparticle into a compact double-lobe pattern whose orientation rotates with axial position. This design yields a clean, well-defined structural evolution across an unprecedented 15 $\upmu$m axial range, with an average rotation rate of 21.8$^\circ/\upmu$m, ensuring robust long-range depth encoding. To recover coordinates from overlapping emitters, we systematically benchmark a conventional matching-pursuit (MP) algorithm against our CNN. While MP performs adequately in sparse scenes, the CNN demonstrates exceptional robustness in dense fields, achieving a 70.24\% localization accuracy at 20 particles per image—substantially outperforming MP's 22.52\% —while also significantly suppressing axial period-jump errors. Experimental validation on nanoparticles confirms reliable single- and multi-particle localization throughout the entire working range. Our results establish ToW-PSFs as a powerful and practical optical encoding strategy for high-density, extended-range 3D localization imaging.
\end{abstract}

\section*{Keywords}
Tornado waves, Point spread function engineering, Three-dimensional localization imaging, Deep learning

\section{Introduction}

Three-dimensional (3D) localization imaging provides a powerful route for resolving nanoscale spatial positions and dynamic trajectories, with broad applications in particle tracking \cite{amselem2023,kashchuk2022,ma2021,nozue2025,zhou2019}, super-resolution microscopy \cite{huang2008,pavani2009,jia2014,gustavsson2018,daly2024,luo2026,von2017}, and volumetric detection \cite{kohler2021,xing2024}. In conventional optical microscopy, however, accurate localization is usually restricted to a narrow depth of field around the focal plane. Moving away from the focus causes the diffraction-limited spot to expand rapidly, reducing the signal-to-noise ratio and localization precision \cite{speidel2003}. Axial scanning can extend the measurable depth range, but sacrifices temporal resolution, making it less suitable for high-speed particle tracking or high-throughput localization tasks \cite{attota2011-2,chen2024}. Multifocal microscopy provides an alternative by simultaneously recording several defocused planes using multiple detection channels \cite{prabhat2007,ram2009}. Although this approach avoids mechanical scanning, it divides detected photons among several imaging planes, thereby limiting the depth range, photon efficiency, and localization precision. These limitations have motivated the use of engineered point spread functions (PSF), which take advantage of additional wavefront-design degrees of freedom to encode axial information into two-dimensional (2D) intensity patterns that are more distinguishable and persistent than the natural defocus evolution of a standard PSF \cite{huang2008,pavani2009,jia2014,nakatani2024,he2019,shechtman2014,shechtman2015}.

Within this framework, representative designs have emerged---including astigmatic PSFs \cite{huang2008}, double-helix (DH) PSFs \cite{pavani2009,nakatani2024}, self-bending PSFs \cite{jia2014,he2019}, and tetrapod PSFs \cite{shechtman2014,shechtman2015}---that encode axial displacement into spot ellipticity, lobe orientation, lateral displacement, or more complex asymmetric intensity patterns. These designs have progressively extended the accessible axial range and improved 3D localization precision compared to standard defocus-based imaging. Among them, rotational PSFs are particularly attractive: they encode depth into the orientation angle of a double-lobe pattern that remains more spatially confined than a naturally defocused spot over an extended axial range \cite{pavani2009,prasad2013}. However, due to the periodicity of the orientation angle, conventional rotational PSFs typically restrict the axial working range to a single rotation period to avoid depth ambiguity. Extending the working distance across multiple rotation periods increases the accessible depth range but introduces multi-period axial ambiguity, where similar lobe orientations correspond to different axial positions. Existing strategies to alleviate this ambiguity---such as asymmetric rotational PSFs \cite{lew2011,bonin2022} and polarization-dependent PSFs \cite{roider2014}---rely on additional intensity or polarization information, but often require more complex PSF structures or additional optical channels. In parallel, deep learning has become a powerful tool in modern optics and computational imaging, with applications in optical metrology \cite{paine2018,goy2018,feng2019,yang2022}, mode recognition \cite{liu2019,giordani2020,wang2021,feng2022,zhou2023,xu2025}, inverse design \cite{peurifoy2018,liu2018-2,jiang2019,nehme2020}, and image reconstruction \cite{sinha2017,nehme2018,rivenson2018,wang2019,xue2019,tseng2021}. In PSF-engineered 3D localization microscopy, convolutional neural network (CNN)-based methods have outperformed conventional matching-pursuit (MP) algorithm in dense-emitter localization and have enabled data-driven PSF optimization \cite{nehme2020,nehme2018}. Given their demonstrated ability to recognize complex optical modes and speckle patterns \cite{wang2021,feng2022}, we use CNN-based decoding to distinguish rotational PSF images from different axial periods, thereby mitigating multi-period axial ambiguity for extended-range 3D localization imaging.

Tornado waves (ToWs), originally introduced as a class of self-rotating structured fields formed by the coherent superposition of vortex circular Airy beams (CABs) with opposite topological charges (TCs) \cite{brimis2020}, provide an attractive physical model for designing rapidly rotating PSFs. Earlier self-rotating fields based on the superposition of high-order Bessel or Laguerre--Gaussian modes can generate rotating lobe patterns during propagation \cite{pavani2009,vasilyeu2009,rop2012,schulze2015,grover2011}, but their rotation rates are limited by the available longitudinal wave-vector difference and the relatively slow phase evolution of the constituent modes. ToWs overcome this limitation by exploiting the self-focusing dynamics \cite{efremidis2010}, enabling faster rotational evolution and large accumulated rotation over extended propagation distances \cite{brimis2020,chen2022}. Subsequent studies have further expanded ToW-related fields through multiring, chirped, and hybrid caustic phase designs \cite{wu2020,jiang2022,wu2024}, demonstrating the rich controllability of their rotational dynamics. These properties make ToWs particularly suitable for constructing depth-sensitive PSFs with stable, compact, and multi-period rotational features. For 3D localization imaging, such a rotating PSF provides a deterministic optical prior that maps particle depth directly to the morphology of a 2D intensity pattern. However, when the rotational evolution spans multiple periods, similar lobe orientations can reappear at different axial positions, making full-pattern decoding essential for resolving the resulting axial ambiguity.

In this work, we present a 3D localization imaging approach based on ToW-PSFs and CNN-based decoding. The ToW-PSF maps the axial position of a particle into a compact double-lobe intensity pattern whose orientation rotates monotonically within the working range. By using a real-space ToW field distribution for phase loading, the compact rotating double-lobe profile enables effective depth encoding over a $15~\upmu\mathrm{m}$ axial range while maintaining a small interlobe distance. We evaluate both a conventional MP algorithm and a CNN-based decoder for recovering particle coordinates from overlapping ToW-PSF images. Numerical and experimental results show that the CNN-based decoder is substantially more robust than MP in dense particle scenes, where sequential fitting and subtraction degrade under severe PSF overlap and axial period-jump errors. Experiments with nanoparticles further validate reliable single-particle and multi-particle localization across the full axial range. These results demonstrate that ToW-PSFs can serve as a compact and depth-sensitive optical encoder for extended-range, high-density 3D localization imaging.

\section{Results and discussion}

\subsection{Fundamental Theory of ToWs}

ToWs are generated by the coherent superposition of two vortex CABs with opposite TCs \cite{brimis2020}. At the initial plane $(z=0)$, the electric field in polar coordinates $(\rho,\theta)$ can be expressed as
\begin{gather}\label{eq1}
    E_{\rm ToW}(\rho, \theta, z=0)=F_1(\rho)\exp(il_1\theta)+F_2(\rho)\exp(il_2\theta),\\
    F_j(\rho)=\mathrm{Ai}\left(\frac{r_j - \rho}{w_j}\right) \exp\left[\frac{a(r_j - \rho)}{w_j}\right],\, (j=1,2)
\end{gather}
where $\mathrm{Ai}(\cdot)$ denotes the Airy function, and $l_1 l_2 < 0$ indicates that the two CABs carry opposite TCs. The parameters $r_j$, $w_j$, and $a$ determine the primary-ring radius, scaling factor, and energy decay factor of each vortex CAB, respectively. The resulting initial intensity distribution appears as a coherent superposition of two hollow, multi-ring beams. During free-space propagation, the self-focusing property of each vortex CAB causes the optical energy to contract sharply toward its focal region. The self-focusing distance of the $j$-th vortex CAB is given by $f_{\mathrm{Ai},j} = \sqrt{w_j^3(r_j+w_j)}$ \cite{papazoglou2011}. After this abrupt self-focusing, the superimposed field evolves into a rapidly rotating optical structure with multiple main lobes. The number of main lobes is equal to the sum of the absolute values of the TCs, $|l_1| + |l_2|$. To achieve maximum rotation in the dual-lobe configuration, we select the lowest-order opposite charges, $l_1=-l_2=1$. The rotational acceleration of the ToW is highly sensitive to the longitudinal overlap of the two vortex CABs, reaching its maximum when their self-focusing planes coincide exactly, i.e., $f_{\mathrm{Ai},1} = f_{\mathrm{Ai},2}$. To satisfy this focal-matching condition while optimizing depth-dependent rotation, the parameters used in this work are $r_1 = 1.0$ mm, $w_1 = 0.0310$ mm, $r_2=0.5$ mm, $w_2=0.0385$ mm, and $a=0.05$ \cite{chen2022}.

\subsection{Experimental Setup and Imaging Model}

\begin{figure}[!b]
  \centering
  \includegraphics[width=0.6\linewidth]{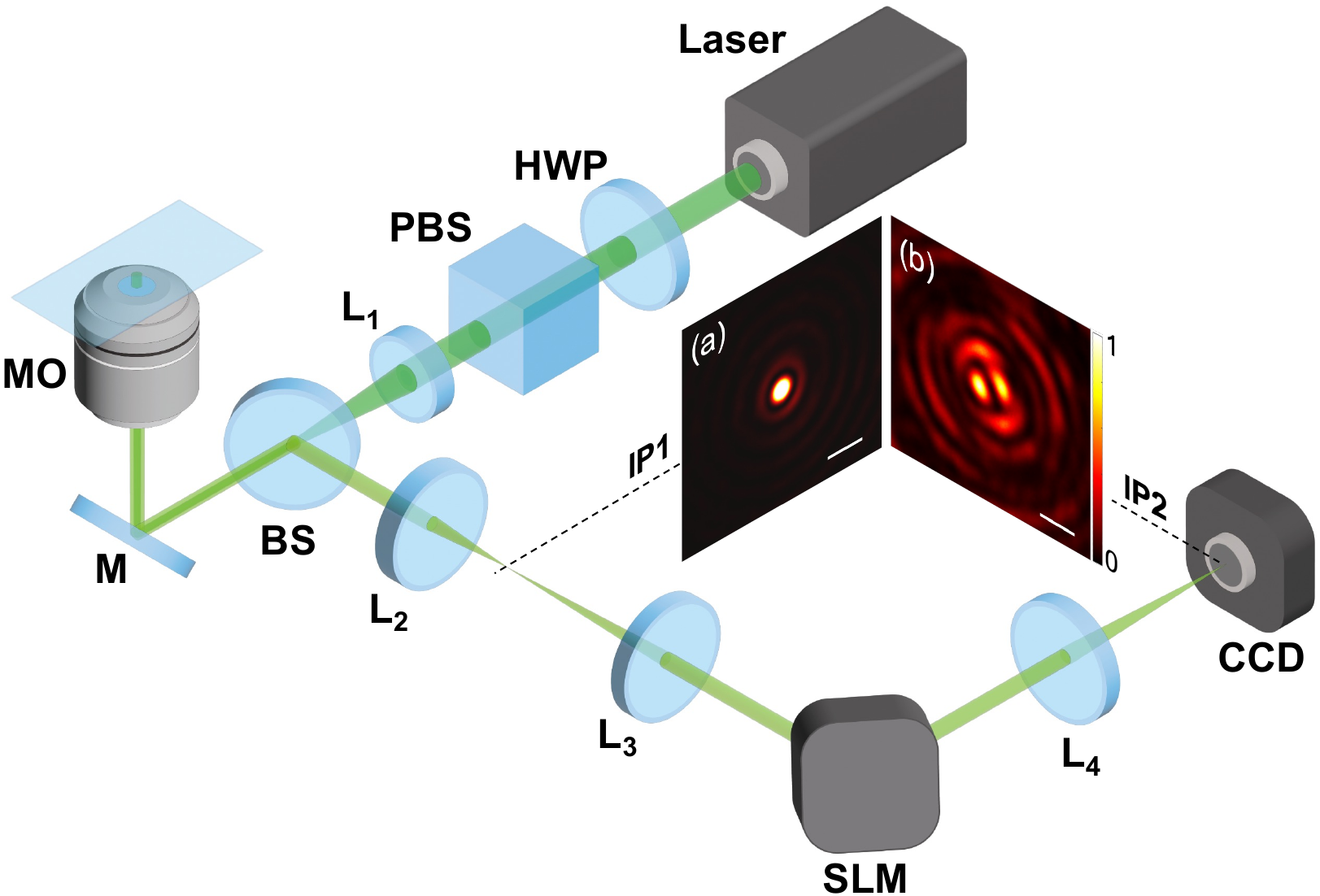}
  \caption{Schematic of the optical setup for 3D localization imaging. Other acronyms include half-wave plate (HWP), polarizing beam splitter (PBS), lens ($L_1,L_2,L_3,L_4$), beam splitter (BS), mirror (M), microscope objective (MO), first image plane (IP1), spatial light modulator (SLM), second image plane (IP2). (a) and (b) Measured intensity distributions of the particle signal at IP1 and IP2, respectively. Scale bars: 1 $\upmu$m.}
  \label{fig1}
\end{figure}

Figure \ref{fig1} shows the optical setup used for 3D nanoparticle localization. The system integrates a microscope imaging system with a $4-f$ modulation module. In the experiments, gold nanoparticles (diameter $d_p = 100$ nm, Sigma-Aldrich) were deposited on a coverslip mounted on a three-axis nanopositioning stage (MLS-1730-3D-CL) to control the particle position $(x_0,y_0,z_0)$. A linearly polarized CW laser (MW-GL-532/2000 mW, $\lambda = 532$ nm) was focused by lens $L_1$ ($f_1 = 225$ mm) and coupled into the microscope objective (MO, Nikon 100$\times$ Oil, NA = 1.25) to illuminate the nanoparticle. Under this illumination, the scattered field from an isolated nanoparticle was treated as an effective point emitter for localization analysis \cite{bohren2008}. The backward scattered light was collected by the MO and lens $L_2$ ($f_2 = 200$ mm) to form an intermediate image at the first image plane (IP1), as illustrated in Figure \ref{fig1}a. The image was then processed by a $4$-$f$ system composed of lenses $L_3$ and $L_4$ ($f_3=f_4=200$ mm) and a spatial light modulator (SLM, Holoeye PLUTO-2-NIR-015, 1920$\times$1080 pixels, each pixel 8 $\upmu$m $\times$ 8 $\upmu$m). The SLM was positioned at the Fourier plane of the $4$-$f$ system, where it loaded a prescribed complex-amplitude modulation function $E_{\mathrm{SLM}}$ through grating-based diffraction encoding \cite{davis1999,bolduc2013}. Finally, the transverse intensity distribution at the second image plane (IP2) was captured by a CCD camera (E3CMOS20000KMA, 5440 $\times$ 3648 pixels, 2.4 $\upmu$m $\times$ 2.4 $\upmu$m pixel area), as illustrated in Figure \ref{fig1}b. In an idealized scenario, if the nanoparticle is located exactly at the nominal object plane, its image at IP1 can be approximated by a $\delta$ function. After passing through $L_3$, this point image is converted into a collimated wave incident on the SLM. The optical field at IP2 is then given by the Fourier transform of the modulation function, $\mathcal{F}[E_{\mathrm{SLM}}]$, where $\mathcal{F}[\cdot]$ denotes the 2D Fourier transform. In practice, however, the field incident on the SLM deviates from the ideal case due to the finite system aperture and the 3D displacement of the nanoparticle. Let $E_p(x,y)$ denote the complex field distribution of the particle image at the SLM plane, the optical field at IP2 can be expressed as $\mathcal{F}[E_pE_{\mathrm{SLM}}]$ \cite{goodman1969}.

The physical principle of 3D localization is that the particle position $(x_0,y_0,z_0)$ is encoded into the complex field distribution $E_p(x,y;x_0,y_0,z_0)$ at the SLM plane. Under the scalar diffraction approximation, the field can be written as \cite{nehme2020,liu2013}
\begin{align}\label{eq3}
  E_p(x,y;x_0,y_0,z_0)=T(x,y)\exp[i\phi_a(x,y;z_0)+i\phi_l(x,y;x_0,y_0)],
\end{align}
where $(x,y)$ are the transverse coordinates at the SLM plane. Here, $T(x,y)$ represents the pupil amplitude limited by the NA of the MO. Due to the finite NA, the pupil acts as a low-pass spatial-frequency filter and can be approximated by a circular aperture function
\begin{align}\label{eq4}
  T(x,y)=
  \begin{cases}
    1, & \sqrt{x^2+y^2}\le r_{\max}, \\
    0, & \sqrt{x^2+y^2}>r_{\max},
  \end{cases}
\end{align}
where $r_{\max}=\mathrm{NA} f_3/M$ is the pupil radius at the Fourier plane, and $M$ is the lateral magnification of the MO. The lateral particle position $(x_0,y_0)$ is encoded as a linear phase tilt
\begin{align}\label{eq4.10}
  \phi_l(x,y;x_0,y_0)=\frac{2\pi M}{\lambda f_3}\left(xx_0+yy_0\right),
\end{align}
which shifts the encoded pattern laterally at IP2. The axial position $z_0$ is encoded through a defocus-dependent phase \cite{nehme2020}
\begin{align}\label{eq4.7}
  \phi_a(x,y;z_0)=z_0 k \sqrt{1-\left(\frac{\mathrm{NA}}{n}\cdot\frac{\sqrt{x^2+y^2}}{r_{\max}}\right)^2},
\end{align}
where $z_0=0$ corresponds to the particle located at the focal plane of the MO, $k=2\pi n/\lambda$ is the wavenumber in the object medium, and $n=1.518$ is the refractive index of the oil immersion medium. This phase modifies the wavefront curvature and, after coupling with the SLM modulation, produces a depth-dependent evolution of the intensity pattern. As a result, the intensity distribution at IP2 carries the 3D position information of the nanoparticle in an optically decodable form. When the modulation function $E_{\mathrm{SLM}}(x,y)$ is loaded onto the SLM, the modulated field from the SLM becomes
\begin{align}
  E(x,y;x_0,y_0,z_0)=E_{\mathrm{SLM}}(x,y)\,E_p(x,y;x_0,y_0,z_0).
\end{align}
After Fourier transformation by lens $L_4$, the field at the second image plane is
\begin{align}\label{eq4.12}
  E_{\mathrm{IP2}}(x,y;x_0,y_0,z_0)
  =&\mathcal{F}\left[E_{\mathrm{SLM}}(x,y)\right]\otimes\mathcal{F}\left[E_p(x,y;x_0,y_0,z_0)\right],
\end{align}
where ``$\otimes$'' denotes 2D convolution. Accordingly, the detected intensity distribution at IP2 defines the PSF of the modulated imaging system
\begin{align}\label{eqPSF}
  \mathrm{PSF}(x,y;x_0,y_0,z_0)=\left|E_{\mathrm{IP2}}(x,y;x_0,y_0,z_0)\right|^2.
\end{align}
This PSF serves as the observable for 3D localization, because its spatial morphology, including main lobes and side lobes, jointly encodes the axial and lateral coordinates of the nanoparticle.

\begin{figure*}[!b]
  \centering
  \includegraphics[width=\linewidth]{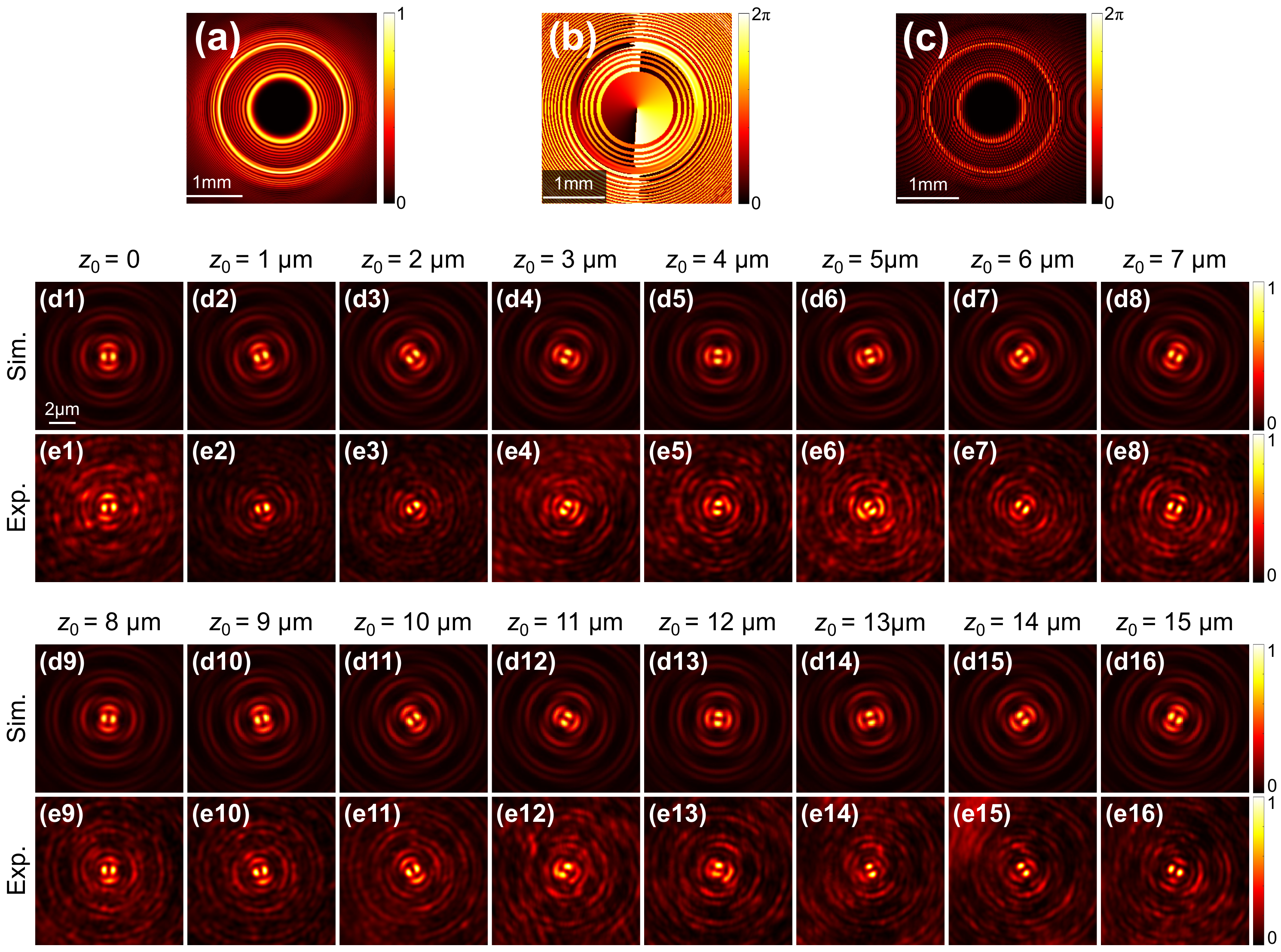}
  \caption{Real-space distribution of the ToW field $E_{\rm ToW}$ used for SLM modulation and the corresponding defocus imaging results. (a) Amplitude distribution. (b) Phase distribution. (c) Phase-only encoding pattern loaded onto the SLM. (d) Simulated evolution of the PSF intensity distributions at IP2 with different axial particle positions $z_0$. (e) Corresponding experimental measurements.}
  \label{fig2}
\end{figure*}

Although the imaging model in Eq. (\ref{eq4.12}) suggests loading the angular spectrum $\mathcal{F}[E_{\mathrm{ToW}}]$ onto the SLM, this scheme places the compact rotating double-lobe feature away from the nominal focal plane because of the initial self-focusing evolution of the ToW field. For practical 3D localization, we therefore adopt the real-space ToW field distribution $E_{\mathrm{ToW}}$ for phase loading. This approach enables axial encoding directly around $z_0=0$. In addition, the hollow intensity distribution of the ToW helps suppress low-spatial-frequency background contributions, such as reflections from the cover glass. A detailed comparison of these two loading strategies is provided in Section 1 of \textcolor{blue}{Supporting Information}. Figures \ref{fig2}a-c show the theoretical amplitude, phase, and phase-only encoding pattern of $E_{\mathrm{ToW}}$, respectively. With this modulation, the system produces a ToW-PSF whose morphology evolves strongly with the particle's axial position. Figure \ref{fig2}d presents the simulated intensity distributions for a particle located at different axial positions $z_0$. Specifically, over the range of $0\le z_0\le 15 \upmu\mathrm{m}$, the field undergoes a pronounced counterclockwise rotation with an average angular rotation rate of $21.8^\circ/\upmu\mathrm{m}$. Here, the angular rotation rate is defined as the derivative of the orientation angle $\varphi$ of the double-lobe PSF with respect to the axial displacement $z_0$, i.e., ${\rm d}{\varphi}/{\rm d}{z_0}$. Because this rotation is deterministic and monotonic over the working range, the axial position of the particle can be quantitatively inferred from the orientation angle of the detected pattern. The corresponding experimental results are shown in Figure \ref{fig2}e. The measured PSFs exhibit a well-defined double-lobe structure, and their rotational evolution with depth agrees well with the theoretical simulations. In the experiment, noticeable background noise appears mainly in the outer-ring region, originating primarily from the reflections of the cover glass. This background field undergoes the same modulation process as the particle signal, producing peripheral noise around the encoded spot. To further highlight the depth-encoding capability of ToW-PSF, we compare its PSF characteristics with the representative DH-PSF design \cite{baranek2014}. The ToW-PSF offers advantages in axial working range, angular rotation rate, and interlobe distance, indicating a more favorable balance between axial working range and PSF compactness. For DH-PSFs with comparable interlobe distance, angular rotation rate, or axial working range, at least one of the other two metrics degrades substantially, whereas the ToW-PSF maintains a compact structure, high angular rotation rate, and extended axial working range simultaneously. Detailed comparisons are provided in Section 2 of \textcolor{blue}{Supporting Information}.

\subsection{Localization Algorithm}

\begin{figure*}[!b]
  \centering
  \includegraphics[width=\linewidth]{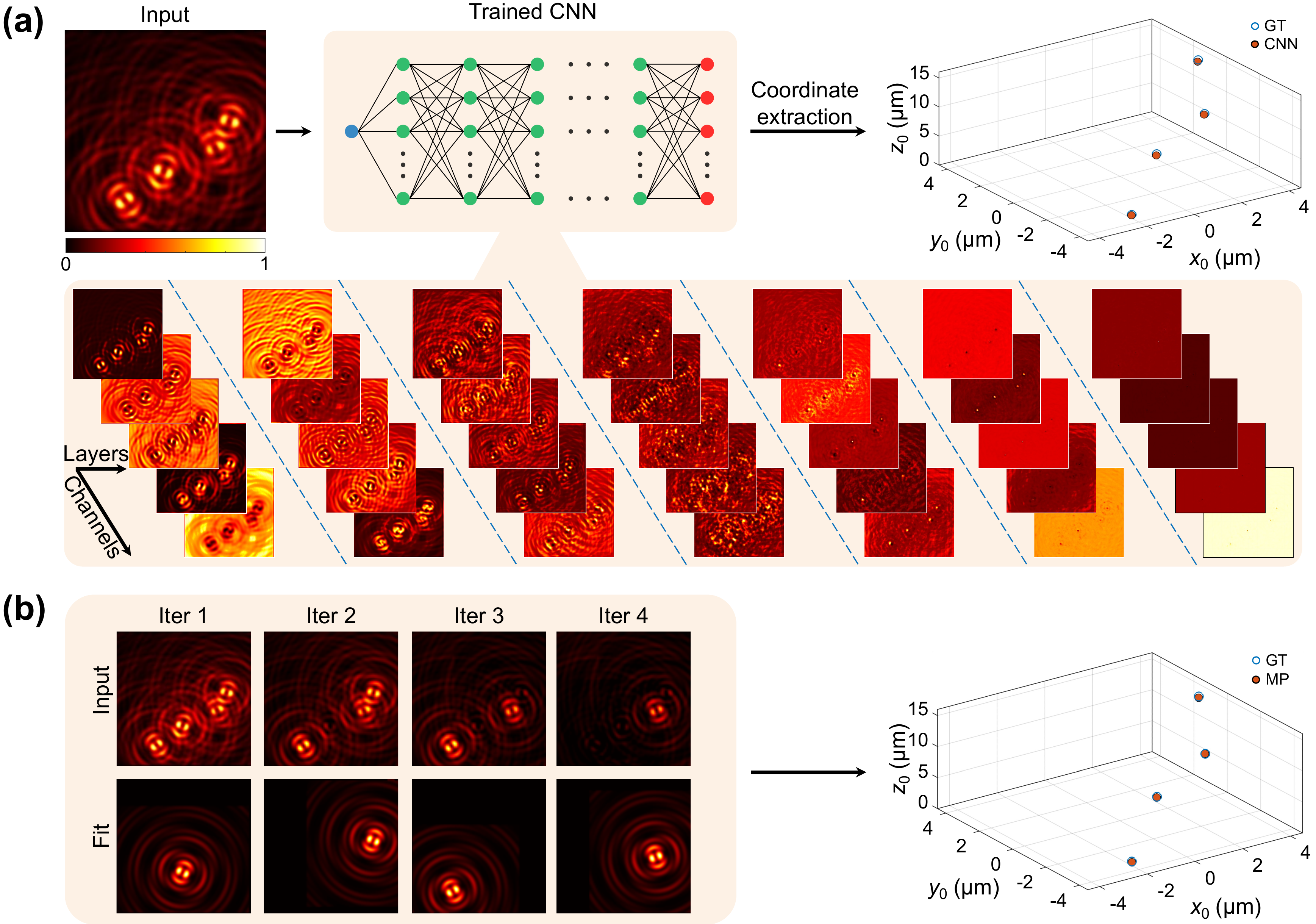}
  \caption{Comparison of CNN and MP workflows for 3D particle localization. (a) CNN localization. The input image containing overlapping PSFs is processed by a trained CNN, which outputs a 3D confidence volume whose voxel values are proportional to the probability of particle presence at the corresponding spatial locations. Particle coordinates are then recovered from this confidence volume by post-processing. The insets below show representative intermediate feature maps from the CNN during inference. The right panel compares the recovered particle coordinates (CNN results, orange dots) with the GT (blue dots). (b) MP localization. The algorithm iteratively identifies, fits, and subtracts the most probable particle signal from the input image until all detectable particles are recovered. The reconstructed particle coordinates (MP results, orange dots) are shown on the right in comparison with the GT (blue dots).}
  \label{fig3}
\end{figure*}

To recover the 3D particle coordinates from the encoded PSFs, we consider both a conventional MP algorithm \cite{mallat1993} and a data-driven CNN-based decoder \cite{nehme2020}, as summarized in Figure \ref{fig3}. For numerical evaluation, the CNN is trained and tested on simulated ToW-PSF images, whereas experimental localization uses a mixed dataset containing simulated images and experimentally synthesized images generated from measured single-particle ToW-PSFs. The trained CNN directly maps the input image to a 3D confidence volume representing the underlying particle distribution in an end-to-end manner. As illustrated by the representative intermediate feature maps in Figure \ref{fig3}a, different feature channels respond to distinct components of the PSF pattern, such as the main lobes and peripheral side lobes. As the network depth increases, these responses become progressively more abstract and integrated. Particle coordinates are then recovered from the predicted confidence volume through post-processing. The recovered particle coordinates agree well with the ground truth (GT) in the example shown in Figure \ref{fig3}a. As shown in Figure \ref{fig3}b, the MP algorithm iteratively identifies, fits, and subtracts the most probable particle signal from the recorded image in a model-based manner until no significant signal remains. Detailed descriptions of both methods are provided in Section 4 of \textcolor{blue}{Supporting Information}.

\subsection{Localization Performance}

\begin{figure*}[!b]
  \centering
  \includegraphics[width=\linewidth]{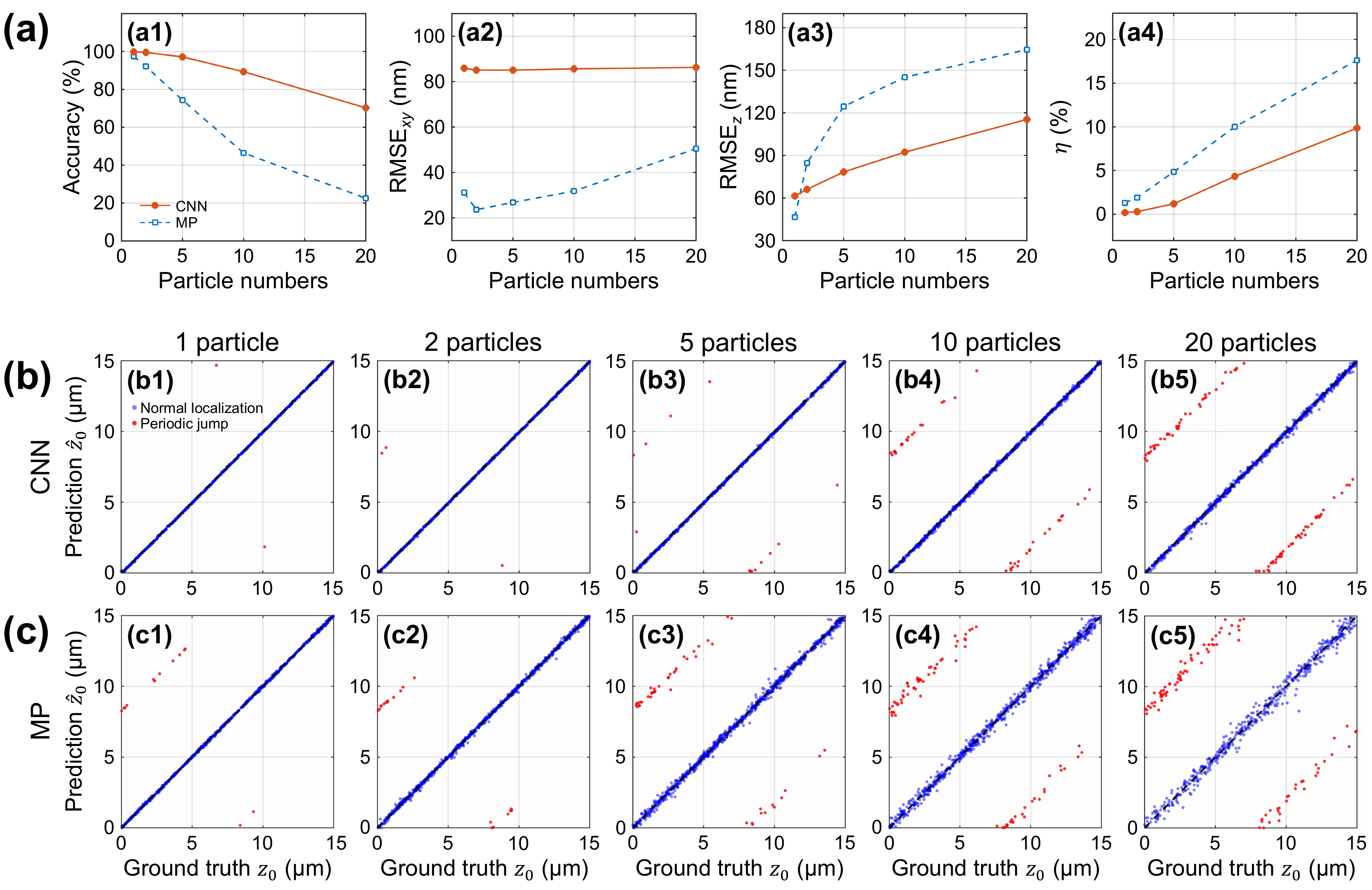}
  \caption{Comparison of localization performance between the CNN and MP algorithms. (a) Quantitative localization metrics as functions of the number of particles per image. Panels (a1)-(a4) show Accuracy, lateral localization error $\mathrm{RMSE}_{xy}$, axial localization error $\mathrm{RMSE}_{z}$, and axial period-jump rate $\eta$, respectively. The Accuracy, $\mathrm{RMSE}_{xy}$, and $\mathrm{RMSE}_{z}$ values in panels (a1)-(a3) were calculated after 3D cylindrical matching and filtering. The axial period-jump rate $\eta$ in panel (A4) was calculated after lateral matching in the $xy$ plane only, so that axial ambiguities caused by the rotational periodicity of the PSF could be evaluated separately. (b) and (c) Scatter plots of the predicted axial coordinate $\hat{z}_0$ with respect to the ground-truth axial coordinate $z_0$. Panel (b) shows the CNN results, and panel (c) shows the MP results. In the ideal case, correctly localized particles lie on the diagonal. Blue points denote axially correct predictions, whereas red points indicate axial period-jump errors. Additional details of the 3D cylindrical matching, lateral matching, and period-jump evaluation criteria are provided in Section 5 of \textcolor{blue}{Supporting Information}.}
  \label{fig4}
\end{figure*}

Although the CNN and MP methods differ fundamentally in their localization mechanisms, the practical significance of this distinction ultimately lies in their performance under particle signal overlap. We therefore next compare the two approaches quantitatively. Figures \ref{fig4}a1-a3 compare their performance as a function of the number of particles per image, evaluated by localization accuracy (Jaccard index) and precision ($\mathrm{RMSE}_{xy}$, $\mathrm{RMSE}_{z}$). For each condition, the total number of ground-truth particles is fixed at 1000. Therefore, for images containing $N_p$ particles, $1000/N_p$ independently generated images are used for evaluation. Detailed metric definitions are provided in Section 5 of \textcolor{blue}{Supporting Information}. In the single-particle case, the two methods show comparable accuracy, with MP yielding better lateral and axial precision. In such sparse scenes, the model-based MP framework can exploit the PSF prior and therefore achieve highly accurate fitting. As particle density increases, the performance of both methods inevitably degrades, yet follows markedly different trends. The localization accuracy of MP decreases rapidly with particle density, dropping to 22.52\% at 20 particles per image, whereas the CNN maintains an accuracy of 70.24\%, nearly three times higher. A similar contrast appears in the precision metrics. Although MP provides higher lateral precision, its $\mathrm{RMSE}_{xy}$ deteriorates quickly as overlap becomes more severe, whereas the CNN exhibits much greater stability.  Axially, MP gives a lower $\mathrm{RMSE}_{z}$ in the single-particle case, whereas the CNN shows lower $\mathrm{RMSE}_{z}$ in multi-particle scenes. These results indicate that the CNN's primary advantage lies not in sparse scenes, where explicit model fitting remains highly effective, but rather in dense scenes. Here, the CNN decodes multiple overlapping PSF features in parallel, making it robust to mutual signal interference.

Beyond general 3D localization performance, the axial error distributions in Figures \ref{fig4}b,c show another aspect of the performance gap. Correctly localized particles lie on the diagonal defined by $\hat{z}_0 = z_0$, and the spread of the identified points around this diagonal reflects the axial localization precision. In addition to this continuous spread, both methods exhibit a second type of error, namely axial misassignment, where the predicted coordinate $\hat{z}_0$ jumps to another axial period due to the 180$^\circ$ rotational periodicity of the PSF. To quantify this effect, we define the axial period-jump rate, $\eta$, as the fraction of particles assigned to another axial period (marked as red points in Figures \ref{fig4}b,c. As quantitatively summarized in Figure \ref{fig4}a4, the CNN yields a substantially lower $\eta$ across all particle densities. In the single-particle case, $\eta$ is only 0.20\% for the CNN, compared with 1.30\% for MP. As the number of particles increases to 20, the corresponding values increase to 9.86\% and 17.61\%, respectively. This difference arises because the sequential identification, fitting, and subtraction steps in MP allow residual overlap and fitting biases to accumulate, leading to error cascading and easily driving the solution toward an incorrect period. The CNN, however, learns multi-scale features from the 2D intensity pattern, including subtle correlations among the main lobes, side lobes, and overlap-dependent structures, which effectively suppresses such ambiguity.

\begin{figure*}[!t]
  \centering
  \includegraphics[width=\linewidth]{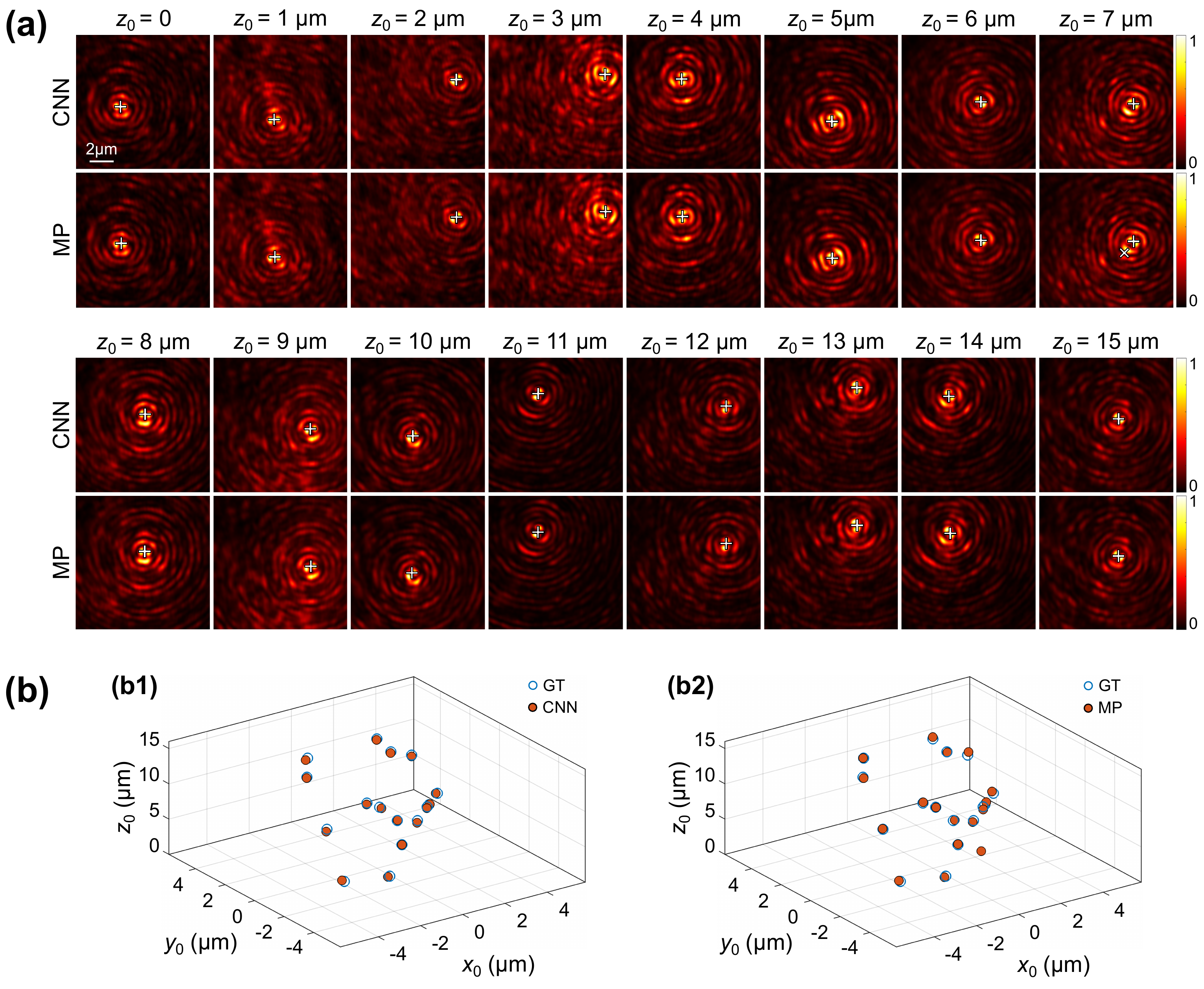}
  \caption{Comparison of single-particle localization results obtained by the CNN and MP across different axial depths and lateral positions. (a) Localization results for particles located at different axial depths $z_0$ and lateral positions $(x_0,y_0)$. The overlaid symbols indicate the lateral coordinates of the predicted particle positions, with correctly localized particles denoted by ``+'' and mislocalized particles denoted by ``$\times$''. (b) Reconstructed 3D particle positions in comparison with the GT, where (b1) and (b2) show the results obtained by the CNN and MP methods, respectively.}
  \label{fig5}
\end{figure*}

To validate the practical viability of the algorithms, we experimentally assess their 3D localization performance, starting with a single-particle case. Over the axial detection range of $z \in [0,15]~\upmu\mathrm{m}$, with measurements sampled at $\Delta z = 1~\upmu\mathrm{m}$, both the CNN and MP methods reliably recover the particle coordinates, as shown in Figure \ref{fig5}. No axial period-jump errors were observed in this single-particle case, indicating that the ToW-PSF retains sufficient depth-encoded structural information across the $15~\upmu\mathrm{m}$ axial range for accurate 3D localization. Under this sparse experimental condition, MP achieves superior lateral precision, with $\mathrm{RMSE}_{xy}=0.05~\upmu\mathrm{m}$, compared to $0.08~\upmu\mathrm{m}$ for the CNN. This shows an advantage of model-based fitting in the absence of signal overlap. Because lateral localization mainly depends on the overall center position of the PSF, which remains relatively stable under moderate distortions and background noise, MP can still provide highly accurate lateral estimates. However, the CNN achieves better axial precision, with $\mathrm{RMSE}_{z}=0.13~\upmu\mathrm{m}$, compared to $0.22~\upmu\mathrm{m}$ for MP. This difference can be attributed to the model mismatch in MP, which relies on a theoretical PSF dictionary that does not capture the optical aberrations and noise present in the experiment. Such imperfections can distort the side lobes and introduce peripheral speckle-like structures. Although these distortions usually have only a limited effect on lateral localization, they can bias the axial estimate because the depth information is encoded in the detailed PSF morphology, including the rotation of the main lobes and the surrounding side-lobe structures. When the side-lobe distortion becomes sufficiently strong, the distorted structure may even be misinterpreted by the algorithm as a valid depth-defining feature. A representative example is observed around $z_0=7~\upmu\mathrm{m}$, where MP shows a noticeable false positive, consistent with a bias induced by distorted experimental side lobes. The CNN alleviates this limitation by learning from the mixed dataset, allowing it to capture the global PSF morphology together with its depth-dependent aberrations and noise characteristics and thereby enabling more robust axial decoding under non-ideal conditions.

\begin{figure*}[!t]
  \centering
  \includegraphics[width=\linewidth]{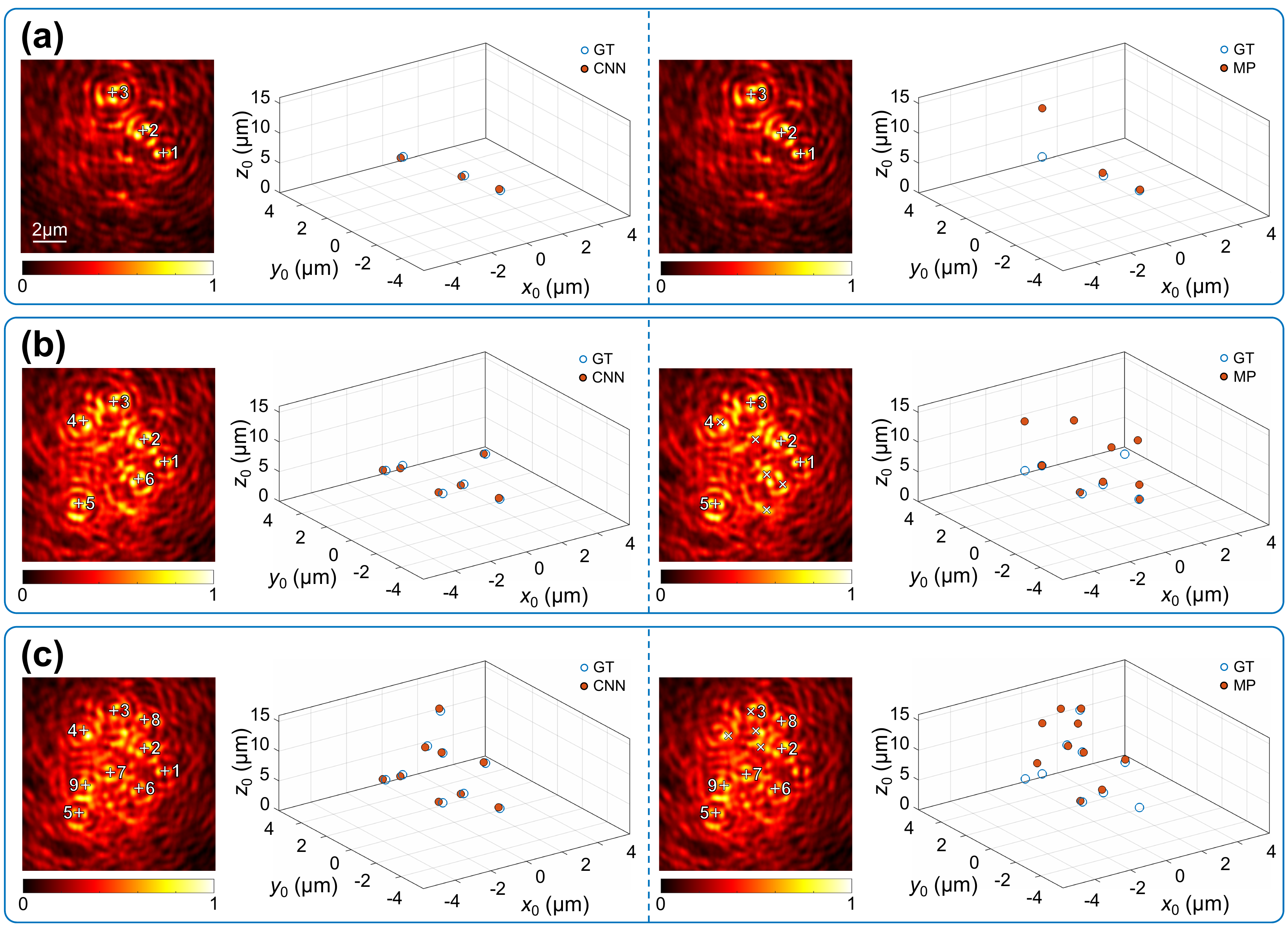}
  \caption{Comparison of synthesized multi-particle localization results obtained by the CNN and MP. (a)-(c) Localization results for images containing 3, 6, and 9 particles, respectively. In each panel, the left and right halves show the results obtained by the CNN and MP, respectively. The overlaid symbols mark the predicted lateral coordinates of the particles, with correctly localized particles denoted by ``+'' and mislocalized particles denoted by ``$\times$''. The corresponding 3D plots compare the reconstructed particle positions with the GT.}
  \label{fig6}
\end{figure*}

Having established the single-particle case, we next evaluate the performance of the CNN and MP algorithms in synthesized multi-particle scenes, where PSF overlap becomes pronounced. These scenes are generated by linearly superimposing independently measured single-particle ToW-PSFs, providing a controlled analogue of incoherent fluorescence multi-emitter imaging while avoiding coherent multi-scatterer interference. As shown in Figure \ref{fig6}, the number of particles per image is progressively increased to examine its effect on 3D localization performance. In the three-particle case (Figure \ref{fig6}a), the CNN identifies all particles, with $\mathrm{RMSE}_{xy}$ and $\mathrm{RMSE}_{z}$ values of 0.11 $\upmu$m and 0.23 $\upmu$m, respectively. Although the MP algorithm retains high lateral precision of 0.03 $\upmu$m, it is less reliable in axial localization. Specifically, particle 3 is misassigned to the next axial period of the PSF, leading to an axial estimate of 12.20 $\upmu$m instead of the true 4.00 $\upmu$m and reducing the overall localization accuracy to 50\%. As the particle number increases to six (Figure \ref{fig6}b), the performance difference becomes more pronounced. The MP algorithm correctly identifies only particles 1, 2, 3, and 5, whereas particle 4 is axially misassigned. Moreover, overlap between particle signals causes side-lobe interference to be interpreted as independent particle signatures, producing both false positives and missed detections. As a result, although the lateral precision remains relatively high ($\mathrm{RMSE}_{xy}=0.09~\upmu\mathrm{m}$), the overall matching accuracy decreases to 36.4\%, with $\mathrm{RMSE}_{z}=0.23~\upmu\mathrm{m}$. Under the same conditions, the CNN correctly identifies all six particles without misidentifications, with $\mathrm{RMSE}_{xy}=0.16~\upmu\mathrm{m}$ and $\mathrm{RMSE}_{z}=0.22~\upmu\mathrm{m}$. When the number of particles is further increased to nine (Figure \ref{fig6}c), the CNN still maintains reliable localization across the 15 $\upmu\mathrm{m}$ axial range, with $\mathrm{RMSE}_{xy}=0.14~\upmu\mathrm{m}$ and $\mathrm{RMSE}_{z}=0.15~\upmu\mathrm{m}$. In comparison, the MP algorithm reaches a matching accuracy of 46.2\%, with $\mathrm{RMSE}_{xy}=0.11~\upmu\mathrm{m}$ and $\mathrm{RMSE}_{z}=0.29~\upmu\mathrm{m}$. The performance gap at higher particle density suggests that the CNN can better handle severe PSF overlap, whereas the sequential identify-fit-subtract procedure of MP becomes more easily affected by interference and error propagation.

\section{Conclusion}

In summary, we have demonstrated a 3D localization imaging approach based on ToW-PSFs, in which the depth-dependent rotation of the ToW field maps particle position into a compact double-lobe intensity pattern over a large usable axial range. Both theoretical analysis and experimental measurements confirm that the ToW-PSF preserves clear structural evolution over a 15 $\upmu\mathrm{m}$ axial interval, providing a robust physical model for long-range depth encoding. To recover 3D particle coordinates from the encoded PSFs, we evaluated both a conventional model-based MP algorithm and a data-driven CNN decoder. While both methods performed reliably in sparse scenes, the CNN showed substantially greater robustness in dense, overlap-dominated conditions, where the MP method became increasingly affected by interference, axial period-jump errors, false positives, and missed detections. These results indicate that combining ToW-based optical encoding with neural-network-based decoding provides an effective means to mitigate multi-period ambiguity and improve localization performance in high-density 3D imaging. More broadly, this work highlights the potential of ToW-PSFs as a practical optical encoder for 3D localization imaging and may find applications in nanoparticle tracking, super-resolution microscopy, and 3D particle sensing.

\section*{Author Contributions}
L.C., R.W., and L.-G.W. designed research; L.C. and R.W. performed research; L.C., R.W., and L.-G.W. developed numerical and analytical tools; L.C., R.W., and L.-G.W. analyzed data; and L.C., R.W., and L.-G.W. wrote the paper. All authors have given approval to the final version of the manuscript.

\section*{Notes}
The authors declare no competing financial interest.

\section*{Acknowledgements}

This work was supported by the National Natural Science Foundation of China (NSFC) (grant Nos. 62375241 and W2521118).

\section*{Supporting information}

The following files are available free of charge.

\begin{itemize}
  \item Comparison of two phase-loading schemes for ToW-PSF generation; comparison between ToW- and DH-PSFs; description of the parallel localization framework for synthesized multi-particle scenes; details of the localization algorithms, including MP implementation, CNN dataset preparation, architecture, training and valida tion procedure, and post-processing; evaluation metrics for localization results; comparison of localization performance between ToW- and DH-PSFs.
\end{itemize}

\bibliography{ref}

\end{document}